# Lattice dynamics and phonon dispersion of van der Waals layered ferromagnet $Fe_3GaTe_2$

Xia Chen, [a] Xi Zhang, *[a] Wenjie He, [a] Yu Li, [b] Jiating Lu, [c] Dinghua Yang, [a] Deren Li, [a] Li Lei,[b] Yong Peng [d] and Gang Xiang *[a]

[a] College of Physics, Sichuan University, Chengdu 610064, China

[b] Institute of Atomic and Molecular Physics, Sichuan University, Chengdu 610064, China

[c] College of Information and Engineering, Sichuan Tourism University, Chengdu 610064, China

[d] School of Materials and Energy, Lanzhou University 730000, China

***Corresponding authors' emails**: xizhang@scu.edu.cn (X.Z.); gxiang@scu.edu.cn (G.X.)

# ABSTRACT

Despite the tremendous progress in spintronic studies of van der Waals (vdW) room-temperature ferromagnet $Fe_3GaTe_2$, much less effort has been spent on its lattice dynamics and possible interaction with spintronic degrees of freedom. In this work, by combining Raman spectroscopy in a wide range of pressure (atmospheric pressure~19.5 GPa) and temperature (80~690 K) with first-principles calculation, we systematically studied the lattice dynamics and phonon dispersion of $Fe_3GaTe_2$. Our results show that the phonon energies of $Fe_3GaTe_2$ located at 126.0 cm$^{-1}$ and 143.5 cm$^{-1}$ originate from the anharmonic $E_{2g}^2$ and harmonic $A_{1g}^1$ vibration modes, respectively. Furthermore, the first room-temperature spin-phonon coupling in vdW ferromagnet is observed with strength of ~0.81 cm$^{-1}$ at 300 K, by identifying Raman anomalies in both phonon energy and full width at half maximum (FWHM) of $E_{2g}^2$ below Curie temperature of $Fe_3GaTe_2$. Our findings are valuable for fundamental and applied studies of vdW materials under variable conditions.

Lattice dynamics is essential for fundamental studies and practical applications of ferromagnetic materials, since the interplay among lattice, charge and spin dynamics significantly influence their intriguing physical properties [1-3]. Very recently, van der Waals (vdW) layered ferromagnet $Fe_3GaTe_2$ has gained extensive attention for its great potential in information storage applications due to its high Curie temperature ($T_c$), large spin polarization and strong perpendicular magnetic anisotropy [4-8]. However, although great progress has been achieved in the spintronic and electronic studies of $Fe_3GaTe_2$ [9-11], much less effort has been spent on the understanding of lattice dynamics and its possible interaction with spintronic and electronic degrees of freedom in $Fe_3GaTe_2$. Despite a few studies on the Raman spectroscopy of $Fe_3GaTe_2$ [5,12], the lattice dynamics of $Fe_3GaTe_2$ is still ambiguous and the assignment of vibrational modes, especially the $E_{2g}^2$ and $A_{1g}^1$ modes, has been based more on intuition than reasoning. As is known, Raman spectroscopy under high pressure up to tens of GPa and high temperature above the $T_c$ can be used to explore atomic-level interactions and advance the understanding of the interplay among lattice vibrations, charge carriers and spin excitations in solid-state materials [13-22]. However, no systematic studies on pressure- and temperature-dependent lattice dynamics of $Fe_3GaTe_2$ have been reported yet.

In this work, we investigate lattice dynamics and phonon dispersion of $Fe_3GaTe_2$ through Raman spectroscopic measurements under a wide range of pressure (atmospheric pressure ~ 19.5 GPa) and temperature (80 K ~ 690 K) in collaboration with first-principles calculations. Based on experimental observation of lattice modes through pressure-dependent Raman measurements and theoretical calculation of phonon energies, two Raman peaks located at 126.0 $cm^{-1}$ and 143.5 $cm^{-1}$ are identified as the $E_{2g}^2$ and $A_{1g}^1$ vibration modes, respectively, which is different from previously-reported results [5,12]. Interestingly, under high pressure, the $E_{2g}^2$ mode softens and the $A_{1g}^1$ mode slightly stiffens, indicating the nature of the $E_{2g}^2$ mode is anharmonic and that of the $A_{1g}^1$ is quasi-harmonic. Furthermore, the temperature-dependent Raman data reveals the spin-phonon coupling in $Fe_3GaTe_2$, where the phonon energy and the FWHM of the $E_{2g}^2$ mode deviate from the anharmonic model as the temperature decreases below the $T_c$ of 360 K.

The microstructural and magnetic properties of bulk $Fe_3GaTe_2$ are first characterized. VdW layered ferromagnet $Fe_3GaTe_2$ composed of Te/$Fe_3$Ga/Te layers exhibits a hexagonal crystal structure, as shown in Figure 1a, belonging to $P6_3/mmc$ (No. 194) space group and $D_{6h}^4$ $(6/mmm)$ point group. Atomic force microscopy (AFM) image in Figure 1b indicates that the studied $Fe_3GaTe_2$ is as thick as ~127 nm. Figures 1c,d present the high-resolution high-angle annular dark-field scanning transmission electron microscopy (HAADF-STEM) images along the [100] and [001] orientations, revealing that the lattice constants $a$ and $c$ are $4.12 \pm 0.03$ Å and $16.27 \pm 0.05$ Å, respectively. Figure 1e shows the energy-dispersive X-ray spectroscopy (EDS) mapping of Fe, Ga and Te atoms, where the nicely arranged atomic distributions underscore high quality of the layered structure of $Fe_3GaTe_2$. For details regarding the XRD pattern of bulk $Fe_3GaTe_2$, one can refer to our prior work [23]. The surface composition and chemical states of $Fe_3GaTe_2$ are characterized by X-ray photoelectron spectroscopy (XPS) of Fe $2p$, Ga $2p$ and Te $3d$ core levels. Figure 1f illustrates the decomposed Fe $2p$ spectrum, where the $2p_{1/2}$ and $2p_{3/2}$ peaks at 720.0 eV and 706.9 eV correspond to the $Fe^0$ state and the peaks at 724.1 eV and 710.5 eV correspond to the $Fe^{3+}$ state, accompanied by the satellite peaks at 727.4 eV and 714.7 eV [6]. Figure 1g shows the peaks at 1144.4 eV and 1117.6 eV, which are identified as Ga $2p_{1/2}$ and $2p_{3/2}$, respectively [6]. Figure 1h illustrates the prominent peaks at 582.7eV and 572.3 eV, which are identified as Te $3d_{3/2}$ and Te $3d_{5/2}$, respectively, and indicate the existence of $Te^{2-}$ [6,24], and the weak peaks at 587.4 eV and 574.1 eV indicate the existence of $Te^{4+}$ due to oxidization [24]. The magnetic properties of bulk $Fe_3GaTe_2$ are examined by superconducting quantum interference device (SQUID) and shown in Figures 1i-1$l$. Figure 1i illustrates the temperature-dependent magnetization curves, indicating that the saturation magnetization ($M_s$) rises from 22.5 emu/g to 54.7 emu/g and the coercivity ($H_c$) increases from 76.6 Oe to 264.2 Oe as the temperature drops from 350 K to 2 K (Figure 1j). Figure 1k shows that the bulk $Fe_3GaTe_2$ exhibits superior $M_s$ under out-of-plane magnetic field at 300 K, revealing its strong perpendicular magnetic anisotropy at room temperature. The field-cooling magnetization curve in Figure 1$l$ reveals that the $T_c$ of bulk $Fe_3GaTe_2$ is~360 K, well above room temperature.

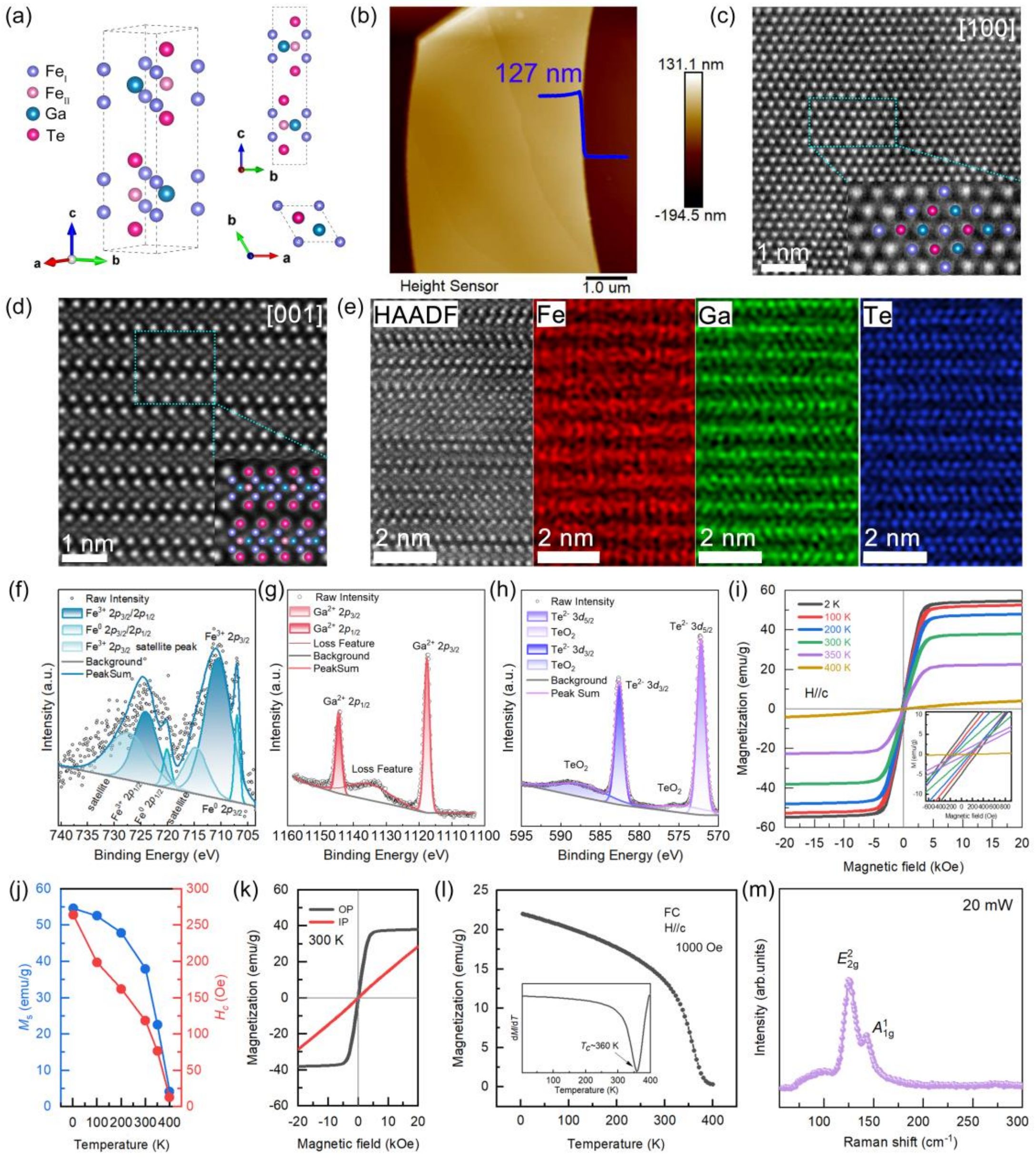

**Figure 1.** (a) Crystal structure of bulk $Fe_3GaTe_2$. (b) AFM image of the 127 nm-thick $Fe_3GaTe_2$ flake. The HAADF-STEM images viewed along the (c) [100] and (d) [001] orientations of the $Fe_3GaTe_2$ flake, where the insets show the magnified atomic structures. (e) STEM-EDS elemental mapping of the $Fe_3GaTe_2$ flake along the [001] orientation. XPS of the fresh $Fe_3GaTe_2$ flake for the (f) Fe, (g) Ga and (h) Te elements. (i) Magnetic field-dependent magnetization curves of bulk $Fe_3GaTe_2$ at various temperatures. The inset shows magnified curves around zero field. (j) Temperature-dependent saturation magnetization and coercivities for bulk $Fe_3GaTe_2$. (k) In-plane and out-of-plane magnetization curves of bulk $Fe_3GaTe_2$ at 300 K. (l) Out-of-plane field cooling magnetization curve of $Fe_3GaTe_2$. The inset presents the temperature-dependent d$M$/d$T$, indicating that the $T_c$ is ~360 K. (m) Raman spectroscopy of $Fe_3GaTe_2$ flake.

According to space group theory, $Fe_3GaTe_2$ owns an irreducible representation of $\Gamma$=4$A_{2u}$+4$E_{1u}$+4$E_{2g}$+4$B_{2g}$+2$A_{1g}$+2$E_{2u}$+2$E_{1g}$+2$B_{1u}$ at the center of the Brillouin zone, where the $E_{1g}$ and $E_{2g}$ modes are doubly degenerate and the $A_{1g}$ and $E_{2g}$ modes are Raman-active phonons, the same as its isostructural counterpart $Fe_3GeTe_2$ [25-27]. At 20 mW excitation power, two phonon energies located at 126.0 cm$^{-1}$ and 143.5 cm$^{-1}$ are observed in the $Fe_3GaTe_2$ flake, as shown in Figure 1m. The first-principles calculations are employed to identify the vibrational modes, as shown in Table I, where the calculated phonon energies of 118.7 cm$^{-1}$ and 142.7 cm$^{-1}$ are assigned to the $E_{2g}^2$ and $A_{1g}^1$ modes, respectively, corresponding to the measured values of 126.0 cm$^{-1}$ and 143.5 cm$^{-1}$. The phonon energies obtained using different computational techniques with different parameters are shown in Table S1 (supporting information), indicating that the choice of parameters in our work results in the calculated results closest to the experimental results. The discrepancy of 7.3 cm$^{-1}$ between the calculated and measured phonon energies of $E_{2g}^2$ is because the nature of the $E_{2g}^2$ mode is *anharmonic*, as we will demonstrate in the following experiments, but the *harmonic* approximation is used in the VASP calculation [28].

The assignment of vibrational modes is further verified by pressure-dependent Raman spectroscopic measurements. The schematic of the measurement setup and the Raman data under various pressures are shown in Figures 2a-2c. Notably, the contour-color map of pressure-dependent Raman intensity in Figure 2d shows that the $E_{2g}^2$ mode at 126.7 $cm^{-1}$ exhibits much higher intensity than the $A_{1g}^1$ mode at 143.8 $cm^{-1}$ does, indicating the microstructural integrity of the $Fe_3GaTe_2$ placed onto the diamond anvil cells (DAC). Fitted by a Gauss+Lor function, the $E_{2g}^2$ phonon energy monotonically decreases from 126.7 $cm^{-1}$ to 106.5 $cm^{-1}$ as the pressure increases from ambient to 7.6 GPa, and then stabilizes, as shown in Figure 2e. Conversely, the phonon energy of the $A_{1g}^1$ mode exhibits an overall blue-shift with some fluctuations: it initially stabilizes at ~143.5 $cm^{-1}$ as the pressure increases from ambient pressure to 7.6 GPa, then further increases to 148.1 $cm^{-1}$ at 17.6 GPa, and finally slightly decreases to 146.5 $cm^{-1}$ at 19.5 GPa, as shown in Figure 2f. Meanwhile, the full widths at half maximum (FWHMs) of $E_{2g}^2$ and $A_{1g}^1$ broaden significantly with increasing pressure, indicating higher phonon scattering and reduced phonon lifetime. It is noted that an inflection occurs at 7.6 GPa for both phonon energies and FWHMs of $E_{2g}^2$ and $A_{1g}^1$, which may be attributed to the isostructural phase transition induced by pressure-driven lattice distortion (Table S2), similar to that in $Fe_3GeTe_2$[17]. Furthermore, the subtle variations in phonon energies and FWHMs between 17.6 GPa and 19.5 GPa may be linked to the non-crystalline of $Fe_3GaTe_2$, as evidenced by the absence of Bragg diffraction peaks in high-pressure XRD data [29]. A more in-depth discussion on the underlying mechanisms governing the variations in phonon energies and FWHMs of $E_{2g}^2$ and $A_{1g}^1$ will be presented in the subsequent sections. Upon releasing pressure, the re-emergence of sharp Raman peak as well as a slight redshift (2 $cm^{-1}$) in the phonon energy of the $E_{2g}^2$ mode under ~0 GPa, as shown in Figures 2c,g , indicate that the $Fe_3GaTe_2$, with gentle responsiveness to external pressure, is suitable for vdW layered flexible sensors [30].

Table I. Calculated and experimental values of phonon energies in bulk $Fe_3GaTe_2$

| Symmetry | Calculated (cm$^{-1}$) NM | FM ground state | | | AFM ground state | Experimental |
|---|---|---|---|---|---|---|
| | D2 | D2 | D3 | Optb86b | Optb86b | 300 K |
| $E_{2g}^1$ | | 23.4 | 17.9 | 16.2 | 20.5 | |
| $E_{1g}^1$ | | 71.4 | 68.7 | 68.3 | 70.3 | |
| $E_{2g}^2$ | 98.3 | 118.7 | 117.4 | 117.1 | 119.9 | 126.0 |
| $A_{1g}^1$ | 151.2 | 142.7 | 144.3 | 141.1 | 139.8 | 143.5 |
| $E_{1g}^2$ | | 206.7 | 210.7 | 210.4 | 209.6 | |
| $E_{2g}^3$ | | 222.5 | 228.0 | 228.4 | 232.1 | |
| $A_{1g}^2$ | | 256.2 | 264.9 | 268.1 | 264.9 | |
| $E_{2g}^4$ | | 328.5 | 323.6 | 325.3 | 326.5 | |

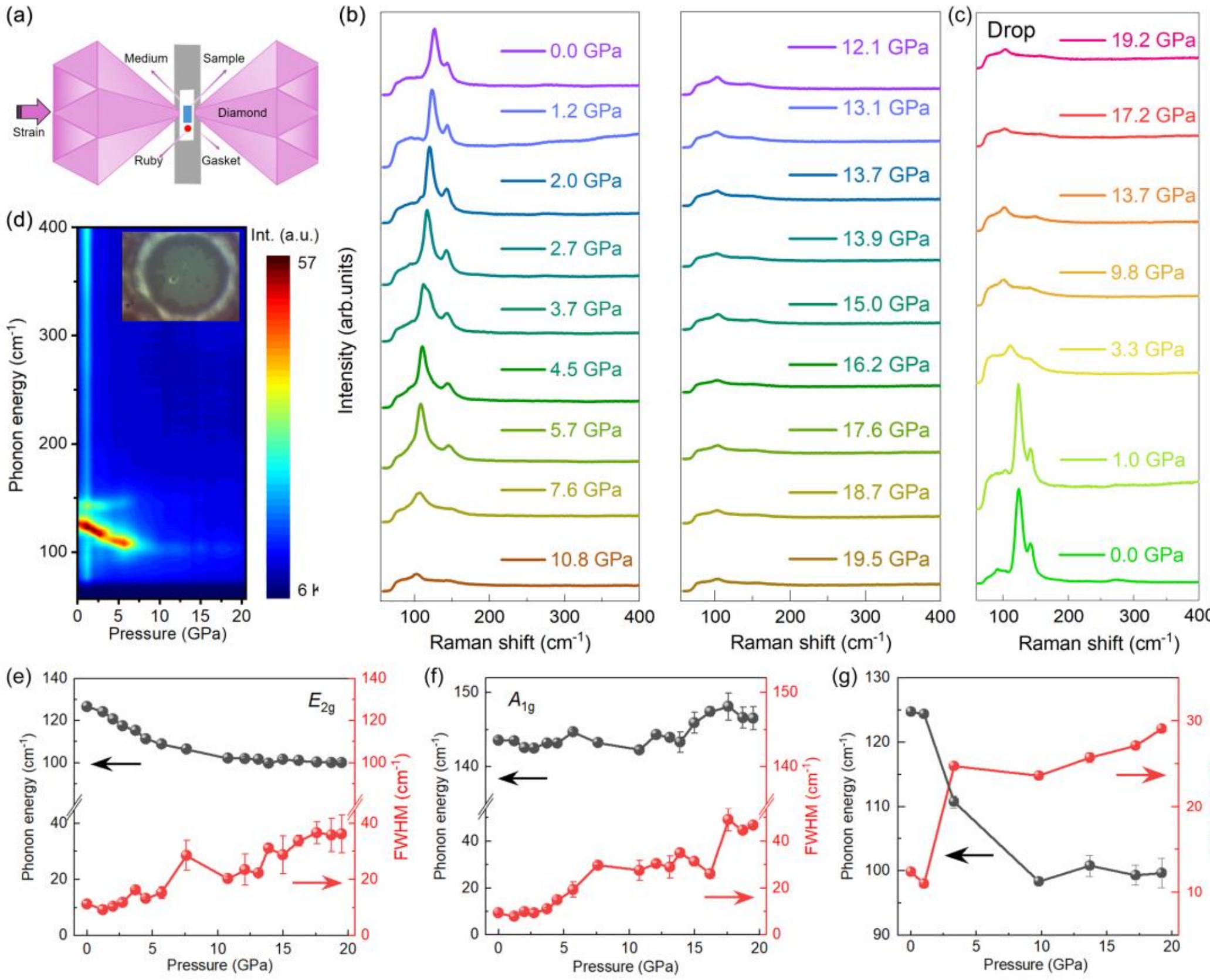

**Figure 2.** *In-situ* high-pressure Raman spectroscopic measurements of $Fe_3GaTe_2$ using DAC. (a) A schematic diagram of DAC. (b) Raman spectroscopies upon applying pressure ranging from 0.0 GPa to 19.5 GPa. (c) Releasing pressure-dependent Raman spectroscopies. (d) Pressures-dependent Raman intensities at 300 K. The inset shows the microphotograph of the $Fe_3GaTe_2$ within DAC. Pressure-dependent phonon energies and FWHMs of the (e) $E_{2g}^2$ and (f) $A_{1g}^1$ modes. (g) Releasing pressure-dependent phonon energies and FWHMs for the $E_{2g}^2$ mode.

Then the vibrational characteristics of the $E_{2g}$ and $A_{1g}$ modes are studied by combining first-principles calculation results and experimental data. The atomic displacement patterns of bulk $Fe_3GaTe_2$ in ferromagnetic state show that the $E_{2g}^2$ mode involves the in-plane atomic vibrations of the Te block ($Fe_1$-Te-$Fe_1$ and $Fe_1$-Te-$Fe_2$) and the Ga block ($Fe_1$-Ga-$Fe_1$ and $Fe_1$-Ga-$Fe_2$), whereas the $A_{1g}^1$ mode involves the out-of-plane atomic vibrations of the Te block (Figure 3a). The phonon dispersion in Figures 3b and Figure S1a-1f (Supporting Information) shows that all frequencies within the Brillouin zone are real, indicating the dynamical stability under high pressure. Meanwhile, the contributions of Fe, Ga and Te atoms to different frequencies are also visualized, where the frequencies below 150 $cm^{-1}$ mainly arise from the Te/Ga atoms and partially stem from the Fe atoms, while higher frequencies are primarily from the Fe atoms, with small contributions from the Te/Ga atoms. Table S2 (Supporting Information) summaries the calculated bond lengths of $d_{\text{Fe1-Te}}$, $d_{\text{Te-Fe1}}$, $d_{\text{Te-Fe2}}$ and $d_{\text{Fe1-Fe1}}$ and bond angles of $\theta_{\text{Fe1-Te-Fe1}}$ and $\theta_{\text{Fe1-Te-Fe2}}$ variation related to the $A_{1g}^1$ mode under various pressures. With increasing pressure, the bond lengths of $d_{\text{Te-Fe1}}$, $d_{\text{Te-Fe2}}$, and $d_{\text{Fe1-Fe1}}$ decrease. Moreover, the lattice constant $c$ and volume decrease significantly with increasing pressure, as shown in Figure 3c. As a result, the inter-atomic force constants are enhanced under pressure[31], leading to an overall blue-shift in the $A_{1g}^1$ mode. The non-monotonic variation on the overall blue-shift background in the $A_{1g}^1$ mode (Figure 2f) may stem from the intricate interplays among pressure-induced changes in isostructural phase transition[17], lattice disorder[30], and electron-phonon[32] or phonon-phonon interaction[33]. By combining X-ray diffraction patterns under high pressure[29] with the calculated pressure-dependent volume in our work, it is found that the value of bulk modulus $K = -V\frac{dP}{dV}$ in $Fe_3GaTe_2$ is positive, where $V$ and $P$ represent volume and pressure, respectively. According to the isothermal model, Grüneisen parameter is given by $\gamma_{iT} = -(\frac{dln(\omega_i)}{dln(V_i)})_T = \frac{K}{\omega_i}(\frac{d\omega_i}{dP})_T$, where $P$ is pressure and $\omega_i$ is the phonon energy of the $i$-th phonon mode [31, 34]. The analysis of the pressure-dependent phonon energy (Figure 2d) suggests that the value of $\gamma_{iT}$ for the $E_{2g}^2$ mode in $Fe_3GaTe_2$ is negative, indicating its anharmonicity, similar to that in $Fe_3GeTe_2$ [31].

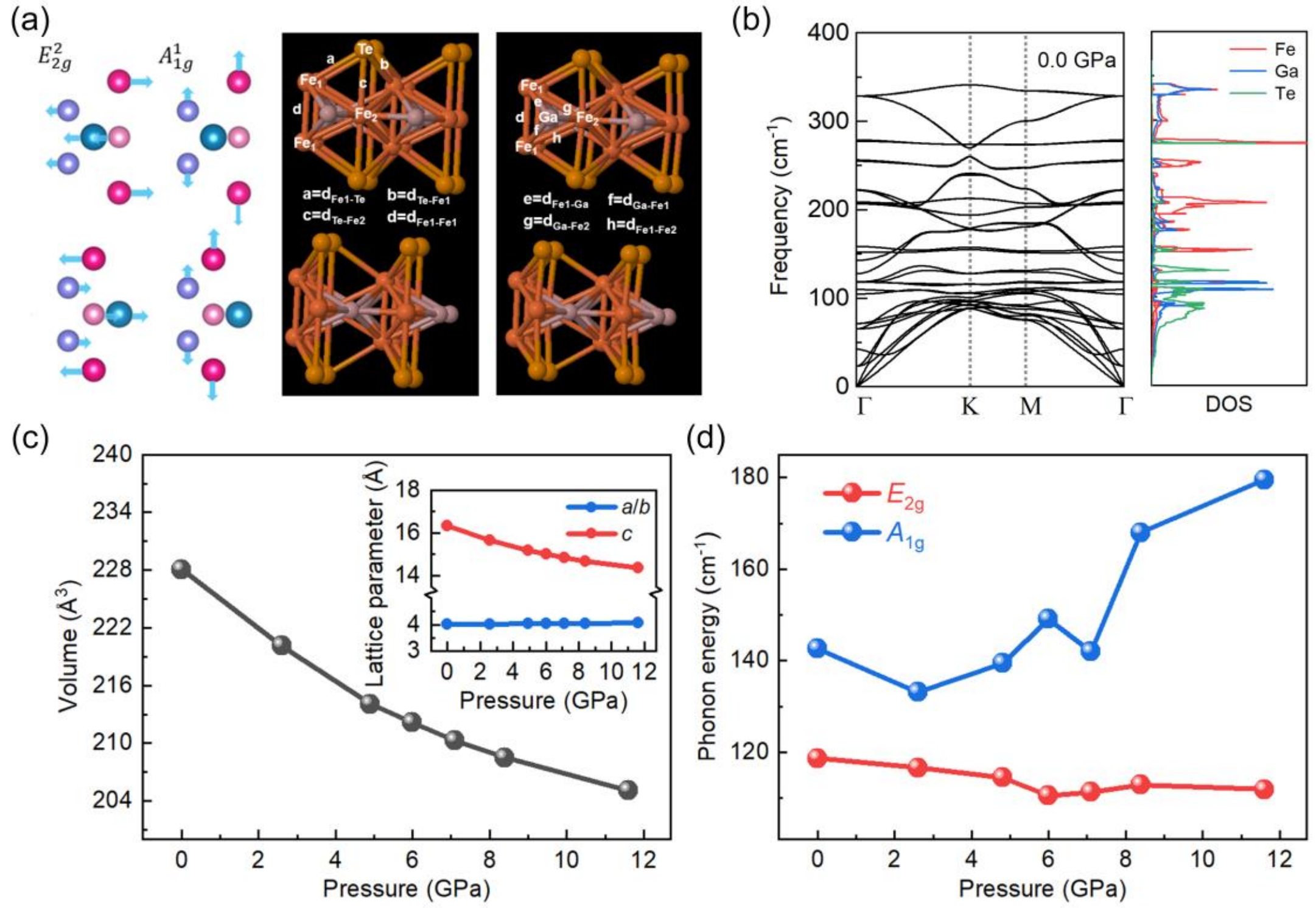

**Figure 3.** (a) Atomic displacement patterns of the bulk $Fe_3GaTe_2$ with ferromagnetic interlayer interaction, where the bright pink, blue, and purple (or pale pink) atoms represent the Te, Ga and Fe atoms, respectively, and the arrows (arrows lengths) represent the vibrational direction (vibrational strength) (left panel). Schematic diagram of changeable atomic structure involving the $E_{2g}^2$ and $A_{1g}^1$ modes under high pressure (right panel). (b) Calculated phonon dispersion and projected phonon density of states (PDOS) of bulk $Fe_3GaTe_2$ under ambient pressure. (c) Calculated pressure-dependent volume of bulk $Fe_3GaTe_2$, where the inset shows the lattice parameters as a function of pressure. (d) Calculated pressure-dependent phonon energies for the $E_{2g}^2$ (red ball) and $A_{1g}^1$ modes (blue ball).

We then discuss the role of spin ordering in the phonon dispersion of $Fe_3GaTe_2$. The temperature-dependent Raman spectroscopic data of $Fe_3GaTe_2$ are shown in Figure 4a. The phonon energies of $E_{2g}^2$ and $A_{1g}^1$ exhibit blueshift as the temperature decreases, as shown in Figures 4b,c. The hardening of $A_{1g}^1$ with decreasing the temperature is the same as that with increasing the pressure, which is normal if one considers that inter-atomic force constants are enhanced in both cases. Interestingly, the hardening of $E_{2g}^2$ with decreasing the temperature is opposite to the softening of $E_{2g}^2$ with increasing the pressure. Given that the vibrational intensity of the Te block is much stronger than that of the Ga block (Figure 3a), the $E_{2g}^2$ phonon energy should mainly originate from the Te block where the bond lengths of $d_{\text{Te-Fe1}}$, $d_{\text{Te-Fe2}}$ and $d_{\text{Fe1-Fe1}}$ decrease with increasing pressure (Table S2, Supporting Information), indicating a hardening tendency. This is opposite to the observed softening of $E_{2g}^2$ with increasing pressure. In order to explain the softening of $E_{2g}^2$, spin ordering has to be considered. Our first-principles calculations reveal that spin ordering plays a crucial role in the $E_{2g}^2$ phonon energy, but not in the $A_{1g}^1$ phonon energy (Table S3, Supporting Information). Under high pressure, the weakening of exchange interactions and magnetic anisotropy [29,31,35] induces decreased spin ordering in $Fe_3GaTe_2$, which is similar to the results observed in $Fe_3GeTe_2$[31], resulting in the decreased spin-correlation function $\langle S_i \cdot S_j \rangle$ and the softening of $E_{2g}^2$. In contrast, with decreasing temperature, the spin ordering could be well-arranged due to suppressed thermal effects and lead to the hardening of $E_{2g}^2$.

Finally, we discuss the spin-phonon coupling effect in the $E_{2g}^2$ mode of $Fe_3GaTe_2$. The phonon energies and FWHMs of $E_{2g}^2$ and $A_{1g}^1$ in Figures 4b-4e can be fitted using phonon anharmonic processes model[36]

$$\omega_{anh}(T) = \omega_0 + A\left[1 + \frac{2}{e^{\frac{\hbar\omega_0}{2k_BT}}-1}\right] + B\left[1 + \frac{3}{e^{\frac{\hbar\omega_0}{3k_BT}}-1} + \frac{3}{\left(e^{\frac{\hbar\omega_0}{3k_BT}}-1\right)^2}\right] \quad (1)$$

$$\Gamma_{anh}(T) = \Gamma_0 + C\left[1 + \frac{2}{e^{\frac{\hbar\omega_0}{2k_BT}}-1}\right] + D\left[1 + \frac{3}{e^{\frac{\hbar\omega_0}{3k_BT}}-1} + \frac{3}{\left(e^{\frac{\hbar\omega_0}{3k_BT}}-1\right)^2}\right] \quad (2)$$

where the $\omega_{anh}$, $\omega_0$, $\Gamma_{anh}$, $\Gamma_0$, $A(B/C/D)$, $\hbar$, and $k_B$ represent modulated phonon energy, phonon energy at 0 K, modulated linewidth, linewidth at 0 K, anharmonic constant, Plank's constant and Boltzmann's constant, respectively. In magnetic materials, the temperature-dependent phonon energies can be described by $\omega(T) \propto \omega(0)+\Delta\omega_{latt}+\Delta\omega_{anh}+\Delta\omega_{sp\text{-}ph}$, proposed by Granado [37]. Here, the $\omega(0)$ represents the phonon energy at 0 K, and the $\Delta\omega_{latt}$, $\Delta\omega_{anh}$ and $\Delta\omega_{sp\text{-}ph}$ correspond to the contributions from the thermal-dependent volume, intrinsic anharmonic effects and the spin-phonon coupling, respectively [37]. For the $A_{1g}^1$ mode, the phonon energy deviates from the anharmonic model below the $T_c$, while the FWHM fits well with the anharmonic model throughout the entire temperature range. Since the FWHM represents the process of phonon decay, it will be influenced by spin-phonon coupling effect instead of lattice parameter change due to the magnetostriction effect [38]. This suggests that the deviation occurring only in the phonon energy of $A_{1g}^1$ may stem from the quasi-harmonic effect due to the magnetostriction-induced structural rearrangements but not from spin-phonon coupling effect in $Fe_3GaTe_2$. Interestingly, both the phonon energy and FWHM of $E_{2g}^2$ well obey with anharmonic model above the $T_c$ of 360 K, confirming the anharmonicity of $E_{2g}^2$, but deviate from the anharmonic model below the $T_c$, revealing the presence of spin-phonon coupling in $Fe_3GaTe_2$, as seen in other materials [28, 39]. The spin-phonon coupling strength of $E_{2g}^2$ can be determined by $\omega \approx \omega_0 + \lambda \langle S_i \bullet S_j \rangle$ [40], where $\omega$ is the energy of the phonon mode, $\omega_0$ is the phonon energy in the absence of the spin-phonon interaction, $\langle S_i \bullet S_j \rangle$ is a statistical average for adjacent spins, and $\lambda$ is spin-phonon coupling strength. We can obtain a saturation magnetization of $Fe_3GaTe_2$ as 37.96 emu/g at 300 K (Figure 1j), corresponding to an average magnetic moment of ~1.12 $\mu_B$ per Fe atom. This yield a $\langle S_i \bullet S_j \rangle$ value of ~0.31 at 300 K. Derived from the experimental data (Figures 4b and 4c), the value of $\Delta\omega=\omega-\omega_0$ of the $E_{2g}^{2}$ mode is 0.25 $cm^{-1}$ at 300 K. Consequently, the spin-phonon coupling strength of $E_{2g}^{2}$ is obtained as ~0.81 $cm^{-1}$ at 300 K (please see more calculation details in Supporting Information). Last but not least, we should point out that the $Fe_3GaTe_2$ flake remains stable throughout the whole process of varying pressure and temperature. In fact, the stability of the $Fe_3GaTe_2$ flake can be distinguished by excitation power-dependent Raman spectroscopic measurements, as shown in Figure S2 (Supporting Information), which provides a fast way for assessing the $Fe_3GaTe_2$ quality.

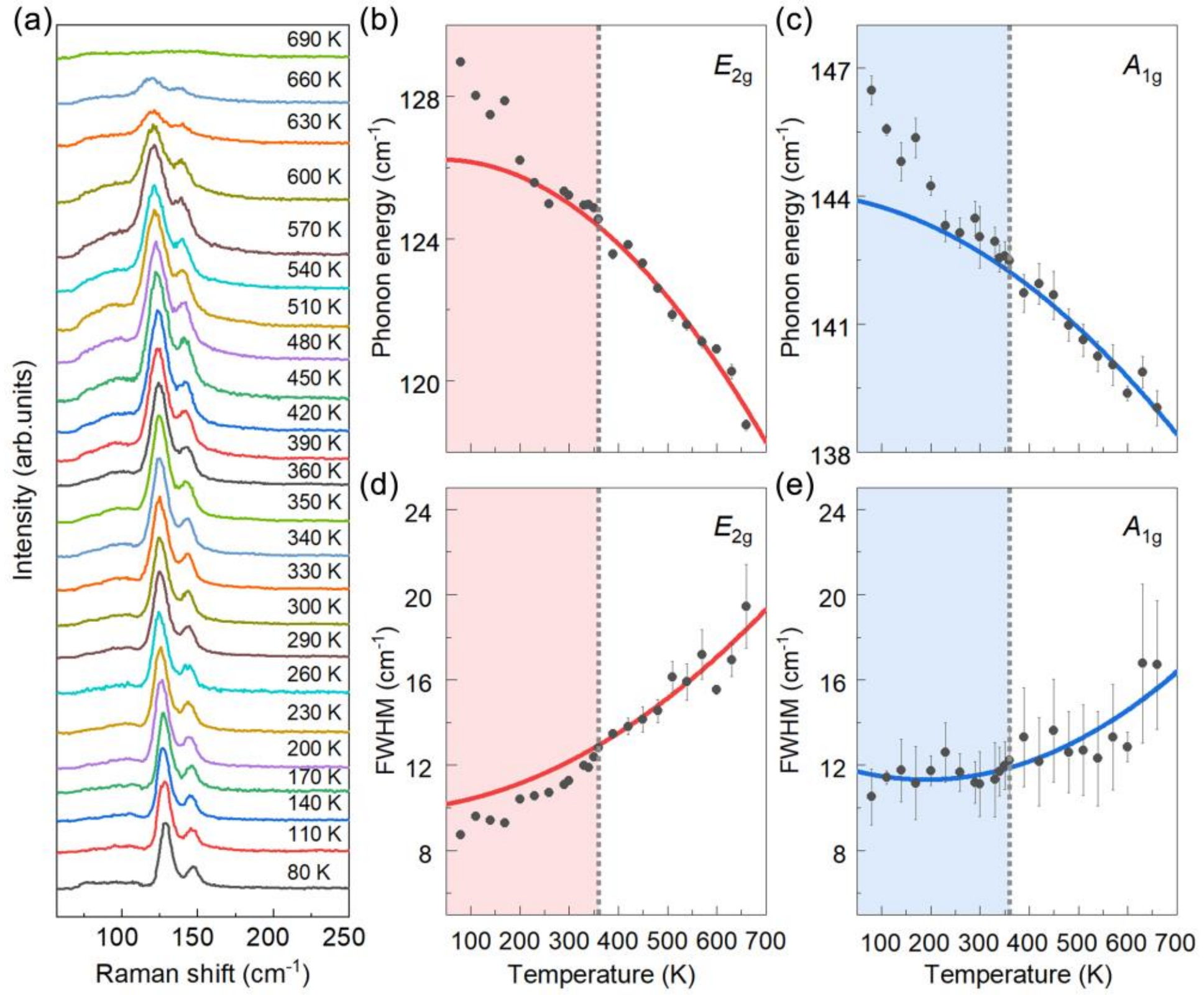

**Figure 4.** (a) Raman spectroscopies of the $Fe_3GaTe_2$ at temperatures varying from 80 K to 690 K. Temperature-dependent phonon energies for the (b) $E_{2g}^{2}$ and (c) $A_{1g}^{1}$ modes. Temperature-dependent FWHMs for the (d) $E_{2g}^{2}$ and (e) $A_{1g}^{1}$ modes. The red and blue lines are fitted results using phonon anharmonic processes.

In conclusion, lattice dynamics, phonon dispersion and spin-phonon coupling of the vdW ferromagnet $Fe_3GaTe_2$ have been thoroughly investigated using Raman spectroscopic measurements and first-principles calculations. Pressure- and temperature-dependent Raman spectroscopic measurements reveal that the vibrational mode of $E_{2g}^{2}$ at 126.0 cm$^{-1}$ is anharmonic and that of $A_{1g}^{1}$ at 143.5 cm$^{-1}$ is quasi-harmonic, and spin ordering plays a crucial role in phonon dispersion of $E_{2g}^{2}$. The anomalous temperature dependency of both phonon energy and FWHM of $E_{2g}^{2}$ below the $T_c$ of 360 K indicates the presence of spin-phonon coupling with strength of ~0.81 cm$^{-1}$ at 300 K in $Fe_3GaTe_2$, underscoring the significant role of spin ordering in lattice dynamics of vdW ferromagnet even at room temperature. In summary, our findings provide a firm groundwork for the development of $Fe_3GaTe_2$ and related vdW materials based spin- and phonon-dependent microelectronic devices.

## ■ METHODS

The $Fe_3GaTe_2$ flakes were mechanically exfoliated and transferred onto $SiO_2$/Si substrate in glove box filled by high-purity $N_2$ (99.999%), and then sealed for physical measurements to avoid oxidization. The microstructures were investigated using HAADF-STEM measurement, where the $Fe_3GaTe_2$ sample was thinned using focused ion-beam milling technique (TESCAN LYRA3 FIB-SEM system). The compositional characterization was performed by X-ray photoelectron spectroscopy (XPS, Kratos AXIS Supra). The thickness was confirmed by atomic force microscopy (AFM, Bruker Dimension Icon). The magnetic properties were measured by superconducting quantum interference device (SQUID, Quantum Design MPMS). Raman spectra were detected in backscattering geometry using a solid-state excitation source of 532 nm (RGB Nova Pro, 300 mW; Laser spot, 5 um). More information on self-assembled optical system for Raman measurement can be found in previous work [41]. For pressure-dependent Raman measurements, we loaded fresh $Fe_3GaTe_2$ flake into a rhenium-gasket diamond anvil cell (DAC) with 500 μm culets, and a mixture of methanol and ethanol with a stoichiometric ratio of 1:4 was used as pressure transmitting medium. Meanwhile, the pressure was calibrated by ruby fluorescence method with error within 0.5 GPa [42,43]. For temperature-dependent measurements, the $Fe_3GaTe_2$ flake was loaded into a heating and freezing microscope stage (THMS600) ranging from 80-690 K. The calculations were carried out utilizing Vienna *Ab initio* Simulation Package (VASP) [44,45], employing projector augmented wave (PAW) [46] pseudopotentials for electron-ion interactions and with the help of spin-polarized generalized gradient approximation (GGA) in the Perdew-Burke-Ernzerhof (PBE) form for exchange-correlation functional [47]. By DFT-D2 approach [48], the interlayer vdW interactions were considered. Initially, the crystal structure was fully releasing with energy cutoff of 450 eV for the plane waves, and total energy (force) convergence criteria was $10^{-8}$ eV ($10^{-6}$ eV/Å). The Brillouin zone was sampled by a $\Gamma$-centered 12×12×2 k-point Monkhorst-Pack mesh. Subsequently, phonon calculations were examined using Finite displacement method in the VASP, followed by analysis with Phonopy software [49,50] to obtain the irreducible representation and corresponding frequencies across the entire Brillouin zone.

## ■ ASSOCIATED CONTENT

Supporting Information

Phonon energies of the $E_{2g}^2$ and $A_{1g}^1$ modes in $Fe_3GaTe_2$ using different calculation methods; Calculated phonon dispersion and projected PDOSs of bulk $Fe_3GaTe_2$ under various pressures; Bond lengths and angles change involved in the $E_{2g}^2$ and $A_{1g}^1$ modes of $Fe_3GaTe_2$ under pressures; Phonon energies of the $E_{2g}^2$ and $A_{1g}^1$modes in $Fe_3GaTe_2$ with different spin ordering; Spin-phonon coupling strength of the $E_{2g}^2$ mode in $Fe_3GaTe_2$; Raman spectroscopy of $Fe_3GaTe_2$ flake at various excitation powers.

## ■ AUTHOR INFORMATION

Corresponding Authors

Xi Zhang—College of Physics, Sichuan University, Chengdu 610065, China; Email:xizhang@scu.edu.cn

Gang Xiang—College of Physics, Sichuan University, Chengdu 610065, China; Email: gxiang@scu.edu,cn

Authors

Xia Chen—College of Physics, Sichuan University, Chengdu 610065, China

Wenjie He—College of Physics, Sichuan University, Chengdu 610065, China

Yu Li—Institute of Atomic and Molecular Physics, Sichuan University, Chengdu 610065,

China

Jiating Lu—College of Information and Engineering, Sichuan Tourism University, Chengdu 610064, China

Dinghua Yang — College of Physics, Sichuan University, Chengdu 610065, China

Deren Li — College of Physics, Sichuan University, Chengdu 610065, China

Li Lei — Institute of Atomic and Molecular Physics, Sichuan University, Chengdu 610065, China

Yong Peng—School of Materials and Energy, Lanzhou University 730000, China

## ■ ACKNOWLEDGMENTS

This work was supported by National Key Research and Development Program of China (MOST) (Grant No. 2022YFA1405100) and National Natural Science Foundation of China (NSFC)(Grant No. 52172272). The XPS test from Analytical and Testing Center in Sichuan University are acknowledged.

## ■ REFERENCES

# Supporting Information for

# Lattice dynamics and phonon dispersion of van der Waals layered ferromagnet $Fe_3GaTe_2$

Xia Chen, [a] Xi Zhang, *[a] Wenjie He, [a] Yu Li, [b] Jiating Lu, [c] Dinghua Yang, [a] Deren Li, [a] Li Lei,[b] Yong Peng [d] and Gang Xiang *[a]

[a] College of Physics, Sichuan University, Chengdu 610064, China

[b] Institute of Atomic and Molecular Physics, Sichuan University, Chengdu 610064, China

[c] College of Information and Engineering, Sichuan Tourism University, Chengdu 610064, China

[d] School of Materials and Energy, Lanzhou University 730000, China

***Corresponding authors' emails**: xizhang@scu.edu.cn (X.Z.); gxiang@scu.edu.cn (G.X.)

**Phonon energies of the $E_{2g}^{2}$ and $A_{1g}^{1}$ modes in $Fe_3GaTe_2$ using different calculation methods**

**Table S1** Phonon energies of $E_{2g}^{2}$ and $A_{1g}^{1}$ modes in $Fe_3GaTe_2$ using different calculation methods

| Various calculation method/parameters | | $E_{2g}^{2}$ (cm$^{-1}$) | $A_{1g}^{1}$ (cm$^{-1}$) |
|---|---|---|---|
| Calculation Method | Finite Displacement | 118.7 | 142.7 |
| | Density Functional Perturbation Theory | 118.2 | 143.8 |
| pseudopotentials | LDA | Imaginary frequencies | |
| | PBE | 118.7 | 142.7 |
| vdW correction | DFT-D2 | 118.7 | 142.7 |
| | DFT-D3 | 117.4 | 144.3 |
| | optB86b | 117.1 | 141.1 |

In order to confirm the rationality of the calculated parameters, we have carried out calculations using different computational techniques with different parameters. The results are shown in Table S1, indicating that the choice of parameters (finite displacement, GGA-PBE and DFT-2) in our work obtains the calculated phonon energies closest to the experimental results. Additionally, the phonon frequencies over the entire Brillouin zone are real and no imaginary frequencies are observed, indicating that the choice of calculation parameters in our work is reasonable.

## Calculated phonon dispersion and projected PDOSs of bulk $Fe_3GaTe_2$ under high pressure

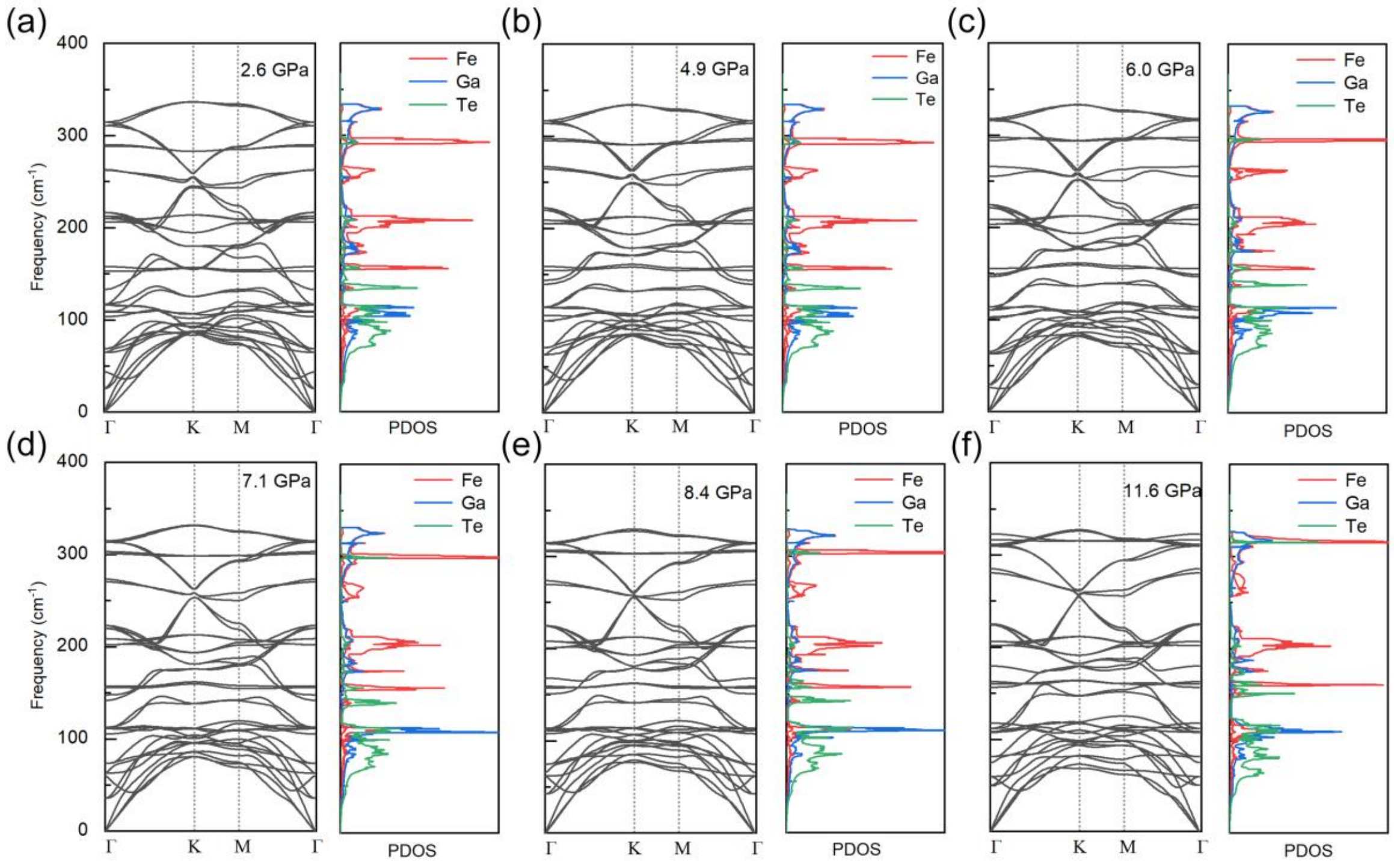

**Figure S1.** Calculated phonon dispersion and projected PDOS of the bulk $Fe_3GaTe_2$ under high pressures of (a) 2.6 GPa, (b) 4.9 GPa, (c) 6.0 GPa, (d) 7.1 GPa, (e) 8.4 GPa, and (f) 11.6 GPa.

Figure S1 shows the pressure-dependent phonon dispersion and projected phonon density of states (PDOSs). All frequencies within the Brillouin zone are real, indicating no structural phase transition under high pressure. Moreover, the contributions of Fe, Ga and Te atoms to different frequencies are also visualized, where the frequencies below 150 $cm^{-1}$ mainly arise from the Te/Ga atoms and partially stem from the Fe atoms, while higher frequencies are primarily from the Fe atoms, with contributions from the Te/Ga atoms.

**Bond lengths and angles change involved in the $A_{1g}^{1}$ and $E_{2g}^{2}$ modes of $Fe_3GaTe_2$ under pressures.**

**Table S2** Bond lengths and angles change involved in the $A_{1g}^{1}$ and $E_{2g}^{2}$ modes of $Fe_3GaTe_2$ under pressures.

| Vibration-al mode | Pressure (GPa) | $\theta_{\text{Fe1-Te-Fe1}}$ (°) | $d_{\text{Fe1-Te}}$ (Å) | $d_{\text{Te-Fe1}}$ (Å) | $\theta_{\text{Fe1-Te-Fe2}}$ (°) | $d_{\text{Te-Fe2}}$ (Å) | $d_{\text{Fe1-Fe1}}$ (Å) |
|---|---|---|---|---|---|---|---|
| $A_{1g}^{1}$ | 0.0 | 56.346 | 4.826 | 2.675 | 73.980 | 2.543 | 2.423 |
| | 2.6 | 56.441 | 4.833 | 2.672 | 74.203 | 2.519 | 2.407 |
| | 4.9 | 56.506 | 4.839 | 2.670 | 74.356 | 2.503 | 2.396 |
| | 6.0 | 56.528 | 4.842 | 2.671 | 74.408 | 2.497 | 2.390 |
| | 7.1 | 56.557 | 4.846 | 2.670 | 74.477 | 2.489 | 2.385 |
| | 8.4 | 56.579 | 4.851 | 2.672 | 74.530 | 2.482 | 2.376 |
| | 11.6 | 56.664 | 4.860 | 2.671 | 74.732 | 2.461 | 2.362 |
| | Pressure (GPa) | $\theta_{\text{Fe1-Ga-Fe1}}$ (°) | $d_{\text{Fe1-Ga}}$ (Å) | $d_{\text{Ga-Fe1}}$ (Å) | $\theta_{\text{Fe1-Ga-Fe2}}$ (°) | $d_{\text{Ga-Fe2}}$ (Å) | $d_{\text{Fe1-Fe2}}$ (Å) |
| $E_{2g}^{2}$ | 0.0 | 55.157 | 2.617 | 2.617 | 152.421 | 2.319 | 4.794 |
| | 2.6 | 54.718 | 2.618 | 2.618 | 152.641 | 2.325 | 4.804 |
| | 4.9 | 54.417 | 2.620 | 2.620 | 152.792 | 2.330 | 4.811 |
| | 6.0 | 54.263 | 2.620 | 2.620 | 152.867 | 2.332 | 4.815 |
| | 7.1 | 54.109 | 2.621 | 2.621 | 152.945 | 2.334 | 4.819 |
| | 8.4 | 53.886 | 2.622 | 2.622 | 153.057 | 2.337 | 4.823 |
| | 11.6 | 53.468 | 2.625 | 2.625 | 153.266 | 2.344 | 4.835 |

**Phonon energies of the $E_{2g}^2$ and $A_{1g}^1$ modes in $Fe_3GaTe_2$ with different spin ordering**

**Table S3.** Phonon energies of the $E_{2g}^2$ and $A_{1g}^1$ modes in $Fe_3GaTe_2$ with different spin ordering. SOC refers to spin-orbit coupling.

| | Spin ordering | $E_{2g}^2$ | $A_{1g}^1$ |
|---|---|---|---|
| Calculation | Non-magnetic | 98.3 | 151.2 |
| | Ferromagnetic+SOC +Out-of-plane-easy axis | 117.7 | 148.5 |
| | Ferromagnetic+SOC +In-plane-easy axis | 117.1 | 149.0 |
| | Ferromagnetic | 118.7 | 142.7 |
| | Antiferromagnetic | 119.9 | 139.8 |
| Experiment | Ferromagnetic | 126.0 | 143.5 |

Table S3 summarizes the phonon energies of the $E_{2g}^2$ and $A_{1g}^1$ modes in $Fe_3GaTe_2$ with different spin ordering. The Table shows that, the $E_{2g}^2$ phonon energy in the magnetic state exhibits much better agreement with experimental result than that in the non-magnetic state, and the $A_{1g}^1$phonon energy either in the magnetic state or in the non-magnetic state is close to the experimental result. The results demonstrate that spin ordering plays a crucial role in the $E_{2g}^2$ phonon energy.

## Note 1. Spin-phonon coupling strength of the $E_{2g}^2$ mode in $Fe_3GaTe_2$

To extract the spin-phonon coupling strength, the relation between the phonon energy and spin-phonon coupling strength can be described using[1-3]

$$\omega \approx \omega_0 + \lambda <S_i \bullet S_j> \tag{S1}$$

where $\omega$ is the phonon energy of the phonon mode, $\omega_0$ is the phonon energy free of the spin-phonon interaction, $<S_i \bullet S_j>$ denotes a statistical average for adjacent spins, and $\lambda$ is spin-phonon coupling strength, which is proportional to $\frac{\partial J}{\partial u}u$ ($J$ is exchange integral and $u$ stands for the ionic displacement of atoms on the exchange path). Deduced from Eq. S1, the spin-phonon coupling strength $\lambda$ is given by

$$\lambda \approx \frac{\omega - \omega_0}{<S_i \bullet S_j>} \tag{S2}$$

According to the relation between magnetic moment and spin quantum number

$$\mu = -g_s \mu_B S = -2\mu_B S$$

where $\mu$ represents the magnetic moment, $g_s$ is Landé g-factor and S represents spin quantum number, the spin-spin correlation can be approximated as[2,3]

$$< S_i \cdot S_j > \approx (\frac{\langle\mu\rangle}{2\mu_B})^2 \tag{S3}$$

The first-principles calculations indicate the average magnetic moment ($<\mu>$) per Fe in $Fe_3GaTe_2$ is 1.94 $\mu_B$ at 0 K, comparable with previous result [4]. Therefore, one can obtain in $Fe_3GaTe_2$ $< S_i \cdot S_j > \approx (\frac{\langle\mu\rangle}{2\mu_B})^2 \approx 0.94$ at 0 K. In addition, the calculated phonon energy of $E_{2g}^2$ in the non-magnetic state is 98.3 $cm^{-1}$ ($\omega_0$), whereas in the ferromagnetic state it shifts to 118.7 $cm^{-1}$ ($\omega$), as shown in Table I. Substituting the values into Eq. S2, the spin-phonon coupling strength of the $E_{2g}^2$ mode is ~21.7 $cm^{-1}$ at 0 K. Experimentally, we have obtained the saturation magnetization of $Fe_3GaTe_2$ as 37.96 emu/g at 300 K (Fig. 1j), namely the average magnetic moment per Fe is ~1.12 $\mu_B$, and then $<S_i \bullet S_j>$ is ~0.31 at 300 K. Derived from the experimental data (Fig. 4b and 4c), the $\Delta\omega = \omega - \omega_0$ of the $E_{2g}^2$ mode is 0.25 $cm^{-1}$ at 300 K, therefore, spin-phonon coupling strength of $E_{2g}^2$ in $Fe_3GaTe_2$ is ~0.81 $cm^{-1}$ at 300 K.

## Excitation power-dependent Raman spectroscopy in $Fe_3GaTe_2$ flake

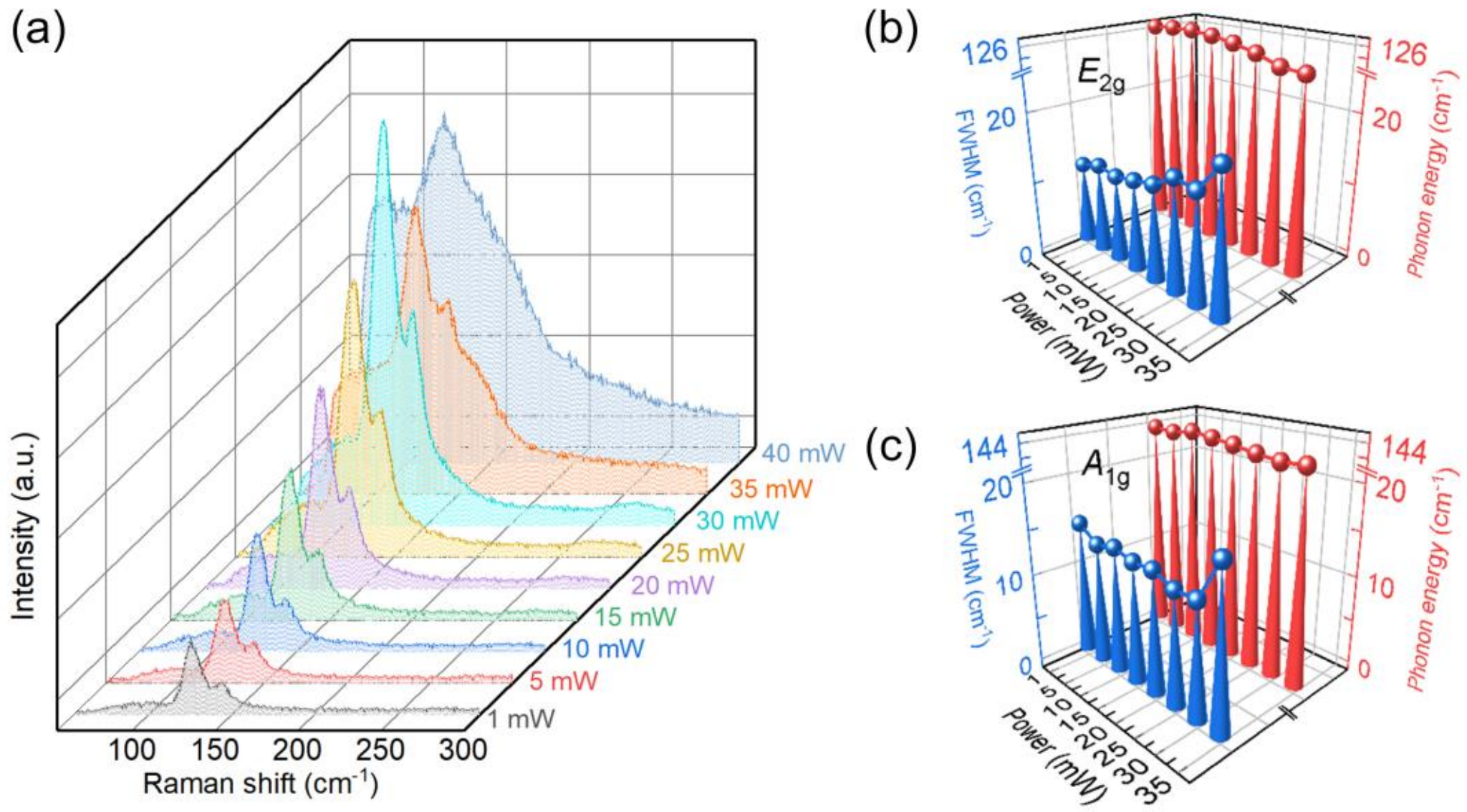

**Figure S2.** (a) Raman spectroscopy of $Fe_3GaTe_2$ flake at various excitation powers. Excitation power-dependent phonon energies and FWHMs for the (b) $E_{2g}^2$ and (c) $A_{1g}^1$ modes.

Excitation power-dependent Raman scattering measurement is performed to evaluate the stability of the $Fe_3GaTe_2$ flake. As shown in Figure S2a, the Raman intensities of the $E_{2g}^2$ and $A_{1g}^1$ modes increase with increasing excitation powers from 1 mW to 20 mW, improving the signal-to-noise ratio. Nevertheless, the $E_{2g}^2$ and $A_{1g}^1$ modes exhibit a redshift above a critical threshold of the excitation power of 20 mW. Furthermore, the emergence of a new Raman peak at an excitation power of 40 mW indicates the structural degradation in $Fe_3GaTe_2$ due to laser radiation (Figures S2b and S2c), which provides a fast way for assessing the $Fe_3GaTe_2$ quality.